# Formal Verification of the Safegcd Implementation [*]

BY RUSSELL O'CONNOR AND ANDREW POELSTRA

Blockstream

*July 22, 2025*

### Abstract

The modular inverse is an essential piece of computation required for elliptic curve operations used for digital signatures in Bitcoin and other applications. A novel approach to the extended Euclidean algorithm has been developed by Bernstein and Yang within the last few years and incorporated into the libsecp256k1 cryptographic library used by Bitcoin. However, novel algorithms introduce new risks of errors. To address this we have completed a computer verified proof of the correctness of (one of) libsecp256k1's modular inverse implementations with the Coq proof assistant using the Verifiable C's implementation of separation logic.

## 0  License

## 1  Introduction

In 2019 Bernstein and Yang published a paper entitled "Fast constant-time gcd computation and modular inversion"[3] which introduced a new greatest common divisor (GCD) algorithm where each iteration of the algorithm depends only on the least significant bit of big integer values and the sign of an auxiliary state variable, called $\delta$.

```python
def gcd(f, g):
    assert f & 1  # require f to be odd
    delta = 1     # additional state variable
    while g != 0:
        assert f & 1  # f will be odd in every iteration
        if delta > 0 and g & 1:
            delta, f, g = 1 - delta, g, (g - f) // 2
        elif g & 1:
            delta, f, g = 1 + delta, f, (g + f) // 2
        else:
            delta, f, g = 1 + delta, f, (g    ) // 2
    return abs(f)
```

**Figure 1.** Python code implementing the high-level safegcd algorithm.[11]

---

[*]. This article has been written using GNU $T_{E}X_{MACS}$ [5].

In each iteration the values $f$ and $g$ are updated such that the $\gcd(f,g)$ remains invariant and $f$ remains odd.[1] When one of the values becomes 0, which must be $g$ since $f$ is always odd, the absolute value of the other value will be the GCD. An analysis of the algorithm shows that for 256-bit input values at most 724 iterations are need.[9]

In order to avoid leaking secret data through side-channels, we want to perform operations in "constant time", which is a technical term that implies no branching and no array look ups based on secret data. Observe that if the loop in Figure 1 did continue running after $g$ reaching 0, the last `else:` branch would be taken and the values of $f$ and $g$ would remain fixed. Therefore, for constant time operation on 256-bit data, we can simply always run the loop 724 times. The `if` branch can also transformed to be constant time using various tricks, such as computing all three outcomes and then performing a "constant time conditional move" using bit masks to select the correct outcome.

GCD algorithms such as this one can be extended to compute modular inverses by adding auxiliary variables to compute Bézout coefficients. This transforms the algorithm into an extended GCD algorithm.

$$u\,f + v\,g = d$$

**Figure 2.** When $d = \gcd(f, g)$, $u$ and $v$ are Bézout coefficients.

In the case where $f$ and $g$ are coprime, then $d = 1$ and $u \equiv f^{-1} \pmod{g}$, i.e. $u$ is the inverse of $f$ modulo $g$.

Bernstein and Yang describe several optimizations of their algorithm which enable high-performance implementations of modular inverse computations on modern computers.

## 1.1 Contributions

This paper describes our formal verification of the correctness of one such real-world, high-performance implementation of this modular inverse algorithm from Bitcoin's libsecp256k1[7] library. To perform this verification, we use Verifiable C[2], which is a framework for verifying the correctness of C code with the Coq proof assistant[14].

# 2  libsecp256k1

libsecp256k1[7] is a C library for handling cryptographic operations over the secp256k1 elliptic curve[4] used in Bitcoin. It served to replace OpenSSL in the Bitcoin codebase.

Modular inverse operations are needed as part of computation of elliptic curve math, which is in turn used for generating and verifying digital signatures in both ECDSA and EC-Schnorr. For signature generation a constant time modular inverse algorithm is needed.

Previously libsecp256k1 used an algorithm based on Fermat's little theorem.[12]

$$a^p \equiv a \pmod{p}$$

**Figure 3.** Fermat's little theorem where $p$ is prime.

---

1. To compute the GCD of two even numbers, simply shift both numbers rightwards together until at least one value becomes odd. However, in our application of computing inverses modulo some large prime $p$, we will always begin with at least one odd input, namely $p$.

This theorem implies that, when $a \not\equiv 0 \pmod{p}$, $a^{p-2} \equiv a^{-1} \pmod{p}$ and modular exponentiation can be used to implement a constant time modular inverse operation. However, such large modular exponentiations are computationally expensive compared to using an extended GCD algorithm.

After safegcd was introduced, Peter Dettman began work on incorporating the new algorithm into libsecp256k1. But before the new algorithm could be accepted, it first needed vetting.

## 3  Formal verification of the safegcd algorithm

It is easy to verify that safegcd algorithm shown in Figure 1 does compute $\gcd(f, g)$ if it terminates. The difficulty lies in proving that the algorithm always terminates. While the safegcd paper offers a termination proof, Bernstein cautioned against using safegcd until the algorithm could be formally verified by a computer proof assistant to ensure it was error free.[1]

Pieter Wuille developed a novel termination argument by computing iterated convex hulls that cover the intermediate values that $f$ and $g$ progress through and showing that eventually the only integer coordinates contained in the hulls have $g = 0$. However, this proof relies on a Python program[9] to compute these convex hulls, which in principle could have bugs. In order to remove doubt, in 2021 we translated the Python code to the Coq proof assistant.[14] We proved that the convex hull algorithm was correct and that it indeed implies the termination of the safegcd algorithm.[10]

This established the correctness of the safegcd algorithm itself, but any real implementation of the safegcd algorithm could still have its own bugs in it.

## 4  Safegcd implementation

Figure 1 presents the most straightforward description of the safegcd algorithm. However, several optimizations are intended to be applied to make the implementation fast. In particular, multiple iterations can be batched together. For example, it is possible to compute an aggregate linear transformation of $f$ and $g$ that performs a batch of 62 iterations at once. This aggregate linear transformation only depends on the lowest 62 bits of $f$ and $g$ and the matrix coefficients each fit within a 64 bit signed value.

There are some other optimizations for computing the Bézout coefficients while avoiding expensive modulo reduction operations. Most of the optimizations used in libsecp256k1's safegcd are described in the "The safegcd implementation in libsecp256k1 explained" documentation.[11] However the documentation does not cover the representation of big integers used in libsecp256k1's modular inverse code. This representation, called signed62, is an array representing values in "base $2^{62}$" using 64-bit signed "digits".

In order to verify the correctness of the libsecp256k1's C code we need a way of formally reasoning about the C programming language. There is one problem however; C does not have a formal language specification.

## 5  Verifiable C

CompCert[6] is a verified C compiler developed in Coq. CompCert provides its own formal semantics of the C programming language, and formally verifies that the resulting assembly code preserves the semantics of this C code.

This doesn't quite guarantee that the compiler is correct. The semantics of C that CompCert defines might not match the official standard, and there have been small errors in the CompCert's formal C semantics in the past. However, it does provide a mostly reliable set of formal semantics of the C language in Coq that we can use for reasoning about C code.

While CompCert provides formal semantics for the C language, it does not come with an effective way of reasoning about C functions. For that we use Verifiable C[2]. Building on top of the formal semantics of C defined by CompCert, Verifiable C implements separation logic for reasoning about C code.

Separation logic[8] is an extension of Hoare logic that adds heap assertions, written as $[a \mapsto v]$, which asserts the value $v$ is stored at address $a$. Additionally, separation logic adds a "separating conjunction", written as the binary operation $A * B$, with the semantics that the heap assertions between expressions $A$ and $B$ are disjoint, i.e. the addresses within expression $A$ do not alias with the addresses within expression $B$.

Specifications of blocks of code are given by Hoare triples $\{P\}C\{Q\}$, where $P$ and $Q$ are the pre- and post-conditions respectively, written in the language of separation logic, and $C$ is a the block of code that will transform any state satisfying $P$ into some state satisfying $Q$.

Let us look at a simple function from libsecp256k1 and its formal specification.

## 5.1 Formally specifying C functions

```
static uint64_t secp256k1_umul128(uint64_t a, uint64_t b, uint64_t* hi)
```

**Figure 4.** `secp256k1_umul128` function signature taken from `int128_struct_impl.h` of libsecp2561 0.6.0.

```
 1 Definition secp256k1_umul128_spec : ident * funspec :=
 2 DECLARE _secp256k1_umul128
 3   WITH a : Z, b : Z, hi : val, sh : share
 4 PRE [ tulong, tulong, tptr tulong ]
 5   PROP(writable_share sh;
 6        0 <= a < Int64.modulus;
 7        0 <= b < Int64.modulus)
 8   PARAMS(Vlong (Int64.repr a); Vlong (Int64.repr b); hi)
 9   SEP(data_at_ sh tulong hi)
10 POST [ tulong ]
11   PROP()
12   RETURN(Vlong (Int64.repr (a * b)))
13   SEP(data_at sh tulong (Vlong (Int64.repr (Z.shiftr (a * b) 64))) hi).
```

**Figure 5.** Specification for `secp256k1_umul128` written in Coq using Verifiable C.

Figure 4 contains the signature of a function from libsecp256k1 for computing the product of two unsigned 64 bit integers that returns the high 64 bits of the result via a passed pointer, and the low 64 bits in the return value. Figure 5 provides a formal specification for this function written using the Verifiable C library.

The specification consists of three parts: a declaration of *mathematical* parameters of the specification, a set of preconditions, and a set of postconditions. You can think of the specification as the statement of a theorem of the form, "for all (mathematical) parameters that satisfy the stated preconditions, if the function arguments are as specified then the return value will be as specified and the stated postcondition will hold afterward". Let us go through this specification line by line.

Line 1 gives a name to this specification. Line 2 say we are giving a specification for the function named `secp256k1_umul128`.

Line 3 declares the mathematical parameters of this specification: integers `a` and `b`, a C value `hi`, and a share value `sh`. The Z you see on this line is the name of Coq's notion of mathematical integers, a.k.a $\mathbb{Z}$, and `a` and `b` are variables of type $\mathbb{Z}$. Keep in mind that the specification's parameters could be completely independent of the parameters of the function that we are providing a specification for.

Lines 4 through 9 are the preconditions given in Verifiable C's syntax. Line 4 also lists the types of `secp256k1_umul128` function's parameters: unsigned long, unsigned long, and a pointer to unsigned long.

Line 5 through 7 give the propositional parts of the specifications preconditions. Line 5 is the requirement that the share `sh` contains the write permission.[2] We will not go into too much detail about shares in this paper as they are mainly used for reasoning about multi-threaded code. Lines 6 and 7 require that the mathematical integers `a` and `b` be in the range $[0, 2^{64})$, i.e. values that fit within the range of C's 64 bit unsigned long type.

Line 8 specifies the `secp256k1_umul128` function's required argument values. The first parameter must be a C long value representing the 64 bit value `a`. Similarly for the second parameter and `b`. The third parameter is the value `hi`.

Line 9 gives the heap part of the specification. It states the data at address `hi` must support the unsigned long type, but otherwise doesn't constrain what that value is. The `sh` parameter defines the read/write permissions that we have for this address. Given line 5, it implies that we have permission to write to (and read from) this address.

Lines 10 through 13 describe the postcondition with line 10 stating the function's return type. We don't have any propositional postconditions in this example, so line 11 is left empty.

Line 12 states the return value of the function, which in this case is the C long value containing the value $a \times b$. Note that the definition of `Int64.repr` converts a mathematical integer to a 64 bit C value by taking the integer value modulo $2^{64}$. Therefore the return value is the lowest 64 bits of this product.

Finally, Line 13 states the heap part of the postcondition. It states that the memory at address `hi` now contains the value of $a \times b \gg 64$, i.e. $a \times b$ divided by $2^{64}$ rounded downward.

# 6 Formally verifying C functions

```
static uint64_t secp256k1_umul128(uint64_t a, uint64_t b, uint64_t* hi) {
    uint64_t ll = (uint64_t)(uint32_t)a * (uint32_t)b;
    uint64_t lh = (uint32_t)a * (b >> 32);
    uint64_t hl = (a >> 32) * (uint32_t)b;
    uint64_t hh = (a >> 32) * (b >> 32);
    uint64_t mid34 = (ll >> 32) + (uint32_t)lh + (uint32_t)hl;
    *hi = hh + (lh >> 32) + (hl >> 32) + (mid34 >> 32);
    return (mid34 << 32) + (uint32_t)ll;
}
```

**Figure 6.** `secp256k1_umul128` function body.

We want to formally prove that the C implementation, given in Figure 6, meets the specification we have given for it. In order to do this we need to bring the C code into Coq somehow, and the tool we use to do that is `clightgen`[3] from the CompCert project.

The `clightgen` program takes the C source code and translates it to a Coq file containing the abstract syntax tree of the C code. This step does also does couple of non-trivial transformations. In particular, `clightgen` explicitly sequences all expressions reading and writing from memory into individual C state-ments, introducing new temporary variables if needed. To prove functions correct, we imports this generated file to access the program's abstract syntax tree.

---

2. In Verifiable C, the write permission always entails read permission as well.
3. Note that `clightgen` is a proprietary program that requires a license to use.

The Coq script used to prove the correctness of `secp256k1_umul128` can be found in Figure 12 from Appendix A. However, this text is simply a transcript of the commands used during interactive proof development, and it is unreadable without the corresponding goal context being manipulated.

When interactively developing a proof in Coq there is a proof state consisting of one or more goal, statements that remain to be proven, and a context of local assumptions that can be used in proofs of the goal. Various tactic commands are given to transform the goal, sometimes solving one goal, other times transforming the goal or splitting it into multiple sub-goals. For example, after the `start_function` tactic, the proof state is as in Figure 7.

```
1 goal
Espec : OracleKind
a, b : Z
hi : val
sh : share
Delta_specs : Maps.PTree.t funspec
Delta := abbreviate : tycontext
SH : writable_share sh
H : 0 <= a < Int64.modulus
H0 : 0 <= b < Int64.modulus
POSTCONDITION := abbreviate : ret_assert
MORE_COMMANDS := abbreviate : statement
______________________________________(1/1)
semax Delta
  (PROP ( )
   LOCAL (temp _a (Vlong (Int64.repr a)); temp _b (Vlong (Int64.repr b));
   temp _hi hi)  SEP (data_at_ sh tulong hi))
  (_ll = ((tulong) (tuint) _a * (tuint) _b);
   MORE_COMMANDS) POSTCONDITION
```

**Figure 7.** Proof state after `start_function` tactic.

All the propositional preconditions that we gave our function's specification appear in the goal's context. These are hypotheses that we are allowed to use when proving the goal. The goal itself contains the value of the local (stack) variables at the start of the function, which have been initialized to the values given in our function's specification. These are `temp _a`, `temp _b`, and `temp _hi`, where the `temp` means a local variable where the address-of operator will not be applied.

Notice the distinction between the C local variable named `_a` and Coq's mathematical variable named `a`. Our Coq variable `a` is a mathematical variable of type $\mathbb{Z}$. It is some arbitrary fixed value. Whereas `_a` is the name we are using for the local C variable. The precondition of our function's specification states that the initial value of this C variable is initialized to the value `Int64.repr a`, but values held by C variables can change during execution.

The goal also contains remaining code to be considered. Only the first statement of the remaining code is displayed, which in this case is the assignment of the product of the values held by `_a` and `_b` to a new variable `_ll`. The rest of the commands are implicitly part of the goal but hidden from being displayed as not to overwhelm the statement. Similarly the postcondition is also there but also hidden from being displayed.

The proof proceeds by *forward reasoning* where we consider the goal's precondition as the "current" abstract state of the program, and then we transform the goal by consuming one statement and updating the "current state" with the strongest postcondition implied by that statement's specification. For function calls, the "next state" is given by the postcondition of the function's specification. For primitive statements, such as assignment, the strongest postcondition is provided by Verifiable C.

```
2 goals
Espec : OracleKind
a, b : Z
hi : val
sh : share
Delta_specs : Maps.PTree.t funspec
SH : writable_share sh
H : 0 <= a < 2 ^ 64
H0 : 0 <= b < 2 ^ 64
PNhi : is_pointer_or_null hi
H1 : field_compatible tulong [] hi
H2 : tc_val' tulong
       (Vlong
          (Int64.repr
             (Z.shiftr a 32 * Z.shiftr b 32
              + Z.shiftr (a mod 2 ^ 32 * Z.shiftr b 32) 32
              + Z.shiftr (Z.shiftr a 32 * (b mod 2 ^ 32)) 32
              + Z.shiftr
                  (Z.shiftr (a mod 2 ^ 32 * (b mod 2 ^ 32)) 32
                   + (a mod 2 ^ 32 * Z.shiftr b 32) mod 2 ^ 32
                   + (Z.shiftr a 32 * (b mod 2 ^ 32)) mod 2 ^ 32) 32)))
______________________________________(1/2)
Vlong
  (Int64.repr
     ((Z.shiftr (a mod 2 ^ 32 * (b mod 2 ^ 32)) 32
       + (a mod 2 ^ 32 * Z.shiftr b 32) mod 2 ^ 32
       + (Z.shiftr a 32 * (b mod 2 ^ 32)) mod 2 ^ 32)
      * 2 ^ 32
      + (a mod 2 ^ 32 * (b mod 2 ^ 32)) mod 2 ^ 32)) =
Vlong (Int64.repr (a * b))
______________________________________(2/2)
data_at sh tulong
  (Vlong
     (Int64.repr
        (Z.shiftr a 32 * Z.shiftr b 32
         + Z.shiftr (a mod 2 ^ 32 * Z.shiftr b 32) 32
         + Z.shiftr (Z.shiftr a 32 * (b mod 2 ^ 32)) 32
         + Z.shiftr
             (Z.shiftr (a mod 2 ^ 32 * (b mod 2 ^ 32)) 32
              + (a mod 2 ^ 32 * Z.shiftr b 32) mod 2 ^ 32
              + (Z.shiftr a 32 * (b mod 2 ^ 32)) mod 2 ^ 32) 32))) hi
|-- data_at sh tulong (Vlong (Int64.repr (Z.shiftr (a * b) 64))) hi
```

**Figure 8.** Proof state after `repeat progressC` tactic.

In our example, `secp256k1_umul128` is a straight line program without loops or branches, and we can immediately consume all the lines of code. After the `repeat progressC` tactic what remains is the two goals shown in Figure 8. The first goal is to prove the value returned in the return statement matches the value stated by the postcondition stated by our function's specification. The second goal is to prove the final heap condition we computed entails the heap condition as stated by the postcondition stated by our function's specification. Both these goals are solved by the usual sort of mathematical reasoning that one does using the Coq proof assistant.

## 6.1 Reasoning about loops

The function `secp256k1_umul128` is a relatively straightforward function to reason about. With more complex functions we cannot always leap directly to a strongest postcondition of the whole function, nor would we necessarily want to even if we could. Proofs usually proceed by considering one or more statements at a time, and then mathematically manipulating the "current state" at that point, simplifying the expression. In addition, loops also need to have an explicit invariant stated for them. Figuring out what those loop invariants are is the essentially hard part of software verification.

For example, consider `secp256k1_modinv64_divsteps_62_var` given in Figure 13 from Appendix B. This function computes a transformation matrix for a batch of 62 steps of the safegcd algorithm. The main loop of that function appears to compute several iterations of the batch at once. First a sequence of zero or more iterations where the value of $g$ remains even the whole time, followed by a batch of at most 6 iterations starting from the case where $g$ is odd.

The first batch is straight forward; $g$ shifted right until it is either odd, or we have reached our maximum of 62 total iterations, and the matrix is updated accordingly. The second batch is a more complicated calculation, but appears to produce an updated $f$ and $g$ and matrix values in accordance with the number of iterations handled.

If the above were true, than the loop invariant would simply be the statement that there exists some value $i$ such that the value of $f$, $g$, and the matrix values are the values given by the safegcd algorithm after $i$ many steps. However, closer examination of the code reveals that the "$g$ is odd" batch is missing the updates for the $u$ and $v$ matrix entries, and missing the bit where $g$ is divided by some power of two. In reality, the code uses the subsequent iteration through the loop to complete those calculations, in combination with the first batch.

Thus we end up needing to specify a quite complex loop invariant shown in Figure 9 that, loosely speaking, states that some $i$ many iterations (at most 62) have been completed except that $g$, $u$, $v$ are all off by a factor of $2^j$ for some $j$ (which is at most $i$).

```
(EX i:nat, EX j:nat, EX u:Z, EX v:Z, EX f:Z, EX g:Z,
 PROP ( Z.of_nat j <= Z.of_nat i <= 62
      ; divstep.Trans.u (divstep.Trans.transN i st) = (u * 2^(Z.of_nat j))%Z
      ; divstep.Trans.v (divstep.Trans.transN i st) = (v * 2^(Z.of_nat j))%Z
      ; eqm (2^(64 - Z.of_nat i)) f (divstep.f (fst (divstep.stepN i st)))
      ; eqm (2^(64 - Z.of_nat i)) g (divstep.g (fst (divstep.stepN i st)))
      )
  LOCAL (temp _i (Vint (Int.repr (62 - Z.of_nat i + Z.of_nat j)));
   temp _g (Vlong (Int64.repr (g * 2^(Z.of_nat j))));
   temp _f (Vlong (Int64.repr f));
   temp _r (Vlong (Int64.repr (divstep.Trans.r (divstep.Trans.transN i st))));
   temp _q (Vlong (Int64.repr (divstep.Trans.q (divstep.Trans.transN i st))));
   temp _v (Vlong (Int64.repr v));
   temp _u (Vlong (Int64.repr u));
   gvars gv;
   temp _eta (Vlong (Int64.repr (divstep.eta (fst (divstep.stepN i st)) + Z.of_nat j)));
   temp _f0 (Vlong (Int64.repr (divstep.f st)));
   temp _g0 (Vlong (Int64.repr (divstep.g st))); temp _t t)
  SEP (data_at_ sh t_secp256k1_modinv64_trans2x2 t;
   debruijn64_array sh_debruijn gv)
)
```

**Figure 9.** Loop invariant used for proving the correctness of `secp256k1_modinv64_divsteps_62_var`.

# 7 Formal verification of the modular inverse C implementation

While the safegcd algorithm was designed to enable a fast constant time implementation, with additional optimizations it also makes for a good variable time implementation, and libsecp256k1 has both constant time and variable time implementations. Furthermore, libsecp256k1 has both 32 bit and 64 bit specific implementations of both of these algorithms.

Our formal correctness proof is specifically for the 64 bit variable time implementation, `secp256k1_modinv64_var`. While this implementation is not directly used in signature generation, it is used after signature generation as sanity check. Therefore, even if the constant time implementation were to have an error in it, it would be caught by this sanity check.

To verify the correctness of `secp256k1_modinv64_var`, the processes of specifying and verifying different C functions was repeated for all functions within its call graph, ultimately specifying and verifying `secp256k1_modinv64_var` itself. Our specification for `secp256k1_modinv64_var` is given in Figure 10.[4]

```
Definition secp256k1_modinv64_var_spec_prime : ident * funspec :=
DECLARE _secp256k1_modinv64_var
  WITH x : Z, m : Z, ptrx : val, modinfo : val, shx : share, sh_modinfo : share,
       sh_debruijn : share, sh_SECP256K1_SIGNED62_ONE : share, gv : globals
  PRE [ tptr t_secp256k1_modinv64_signed62
      , tptr t_secp256k1_modinv64_modinfo
      ]
    PROP( Z.Odd m
        ; 0 <= x < m
        ; m < 2^256
        ; prime m
        ; writable_share shx
        ; readable_share sh_modinfo
        ; readable_share sh_debruijn
        ; readable_share sh_SECP256K1_SIGNED62_ONE
        )
    PARAMS(ptrx; modinfo)
    GLOBALS(gv)
    SEP(data_at shx t_secp256k1_modinv64_signed62
          (map Vlong (Signed62.reprn 5 x)) ptrx;
        data_at sh_modinfo t_secp256k1_modinv64_modinfo (make_modinfo m) modinfo;
        debruijn64_array sh_debruijn gv;
        data_at sh_SECP256K1_SIGNED62_ONE t_secp256k1_modinv64_signed62
          (map Vlong (Signed62.reprn 5 1)) (gv _SECP256K1_SIGNED62_ONE))
  POST [ tvoid ]
    PROP()
    RETURN()
    SEP(data_at shx t_secp256k1_modinv64_signed62
          (map Vlong (Signed62.reprn 5 (modInv x m))) ptrx;
        data_at sh_modinfo t_secp256k1_modinv64_modinfo (make_modinfo m) modinfo;
        debruijn64_array sh_debruijn gv;
        data_at sh_SECP256K1_SIGNED62_ONE t_secp256k1_modinv64_signed62
         (map Vlong (Signed62.reprn 5 1)) (gv _SECP256K1_SIGNED62_ONE)).
```

**Figure 10.** Specification for `secp256k1_modinv64_var` written in Coq using Verifiable C.

The function has two parameters: `ptrx` is a pointer to a representation the value whose modular inverse we want to compute, expressed as an array in libsecp256k1's signed62 format, and `modinfo` is a pointer to a `secp256k1_modinv64_modinfo` structure that contains the modulus along with some auxiliary data.

The specification is additionally parameterized by mathematical integers m and x, representing the modulus and the value whose modular inverse we wish to compute, several shares, which provide suitable read and write permissions to the various pointers, and a collection of global variables.

---

[4]. We actually have two formal specifications of `secp256k1_modinv64_var`, a more complex specification only requiring x and m to be coprime (or x to be 0) and the one given in Figure 10 which requires m to be prime. The later is derived from the former and we expect the later to be easier to apply in practice.

The propositional preconditions are the modulus m is odd and prime and less than $2^{256}$, and the input x is between 0 and m. The memory addressed by ptrx must be writable (and readable), and the other memory needs to be readable.

The heap precondition requires that the array pointed to by ptrx must be the canonical representation of the integer value x expressed in signed62 format. The structure pointed to by modinv must be the canonical representation of the value m and its associated data. Also the global variables (which are actually global constants) for the debruijn array and SECP256K1_SIGNED62_ONE must still be at their initial values.

The heap postcondition are nearly identical to the heap preconditions, implying that the functions leave the heap in the state it began with, with the exception that the array pointed to by ptrx must now be the canonical representation of the modular inverse of x expressed in signed62 format.[5]

The formal Coq proof that the implementation of secp256k1_modinv64_var meets this specification is given in body_secp256k1_modinv64_var in verif_modinv64_impl.v which can be found in our GitHub repository.[13]

## 7.1 A close call

Were any implementation errors found during our verification efforts? No, the implementation was correct. However, there was one instance of one might call a "close call".

In Section 4 of the "The safegcd implementation in libsecp256k1 explained" documentation[11], when talking about a transformation matrix $t$ where $t = \begin{pmatrix} u & v \\ q & r \end{pmatrix}$ used to batch $N$ iterations of the safegcd algorithm, the guide notes:

> It can be shown that $|u|+|v|$ and $|q|+|r|$ never exceed $2^N$ [...], and thus a multiplication with $t$ will have outputs whose absolute values are at most $2^N$ times the maximum absolute input value.

Because of the subsequent division by $2^N$, this then implies that after each batch of $N$ iterations the magnitude of the new values of $f$ and $g$ can never exceed in the maximum magnitude of the previous values of $f$ and $g$. However, there is one optimization in the C code that isn't mentioned in the documentation which is shown in in Figure 11.

```
/* Determine if len>1 and limb (len-1) of both f and g is 0 or -1. */
fn = f.v[len - 1];
gn = g.v[len - 1];
cond = ((int64_t)len - 2) >> 63;
cond |= fn ^ (fn >> 63);
cond |= gn ^ (gn >> 63);
/* If so, reduce length,propagating the sign of f and g's top limb
   into the one below. */
if (cond == 0) {
    f.v[len - 2] |= (uint64_t)fn << 62;
    g.v[len - 2] |= (uint64_t)gn << 62;
    --len;
}
```

**Figure 11.** Optimizing the signed62 array size of $f$ and $g$ within the implementation of secp256k1_modinv64_var.

The signed62 arrays containing the 256-bit values $f$ and $g$ need to have 5 entries. However, because the absolute value of $f$ and $g$ generally shrinks during as the algorithm progresses, we can shrink the length of these arrays as their values become small enough. This allows the function to spend slightly less time looping through them.

Perhaps the reader can already spot the potential issue. Similar to two's complement, the most negative value that can be represented in array of length $L$ is $-2^{63+62L}$, but the most positive value than can be (canonically) represented is $2^{63+62L} - 1$. Observe that the most negative value doesn't have a corresponding positive value of the same magnitude. However, the only constraint we mentioned so far on subsequent values of $f$ and $g$ is that they aren't larger in magnitude. Thus, in the unlikely event that during an intermediate iteration where $g$ is exactly $-2^{63+62L}$ and $|f| < |g|$, then according to what we know so far, on the next iteration $g$ could be as large as $2^{63+62L}$. This next value of $g$ is no larger in magnitude than the previous value of $g$, yet it is beyond the range of values supported by an array of length $L$. This would result in an unhandled overflow, and the function would produce some sort of erroneous result.

---

5. modInv is defined in such a way that `modInv 0 m = 0`. Hence, for a 0 valued input, a 0 valued output must be returned.

The above is essentially the nightmare scenario that we worry about; an obscure unhandled corner case that is essentially impossible to hit by randomized testing, but could perhaps be exploited by an adversary reverse engineering an input to cause this failure to occur. A corner case that can only really be found by performing formal verification.

Fortunately, a more refined analysis of possible transformation matrices shows that additionally $-2^N < u + v$ and $-2^N < q + r$. From this we were able to conclude that it is impossible for $g$ be negated between iterations, and therefore the nightmare scenario we described above cannot occur.

## 8 Verifiable C's limitations

Verifiable C comes with certain limitations. As mentioned earlier, there isn't a formal specification of the C language, and in lieu of that we are using for formal semantics of the C language that CompCert defines. This specification includes aspects that are not strictly defined in the C language, such as two's complement signed integers. More importantly, CompCert specifies its own order of execution of memory operations when evaluating function arguments by translating programs to ones in which memory operations are always explicitly sequenced.

Some of these limitations can be addressed. For instance, one could use the CompCert translated program when compiling with other C compilers to guarantee the same order of memory operations. Furthermore, Verifiable C's framework is designed keep signed integer overflow and underflow undefined, even though CompCert semantics defines that behaviour.

Perhaps, the most important limitation is that Verifiable C specifications only support what is known as *partial correctness*, which means that the postcondition holds *if* the function terminates, but the specification itself does not guarantee termination. Thus we have only proven that if `secp256k1_modinv64_var` terminates, it yields a correct answer, but technically we have not proven that it always terminates.

However it is clear from a combination of our proof and inspecting the code the it must terminate. There is an assertion in `secp256k1_modinv64_var` that fails if the main loop counter exceeds 12, and our formal proof does verify that this, and all other assertions, never fail. Other loops in other subroutines also be inspect by hand to see they they all terminate as well.

Another limitation of Verifiable C is that it does not handle passing or returning of C structures by value, nor does it handle assignment of C structures. Fortunately libsecp256k1 happens to be written in a style where structures are always passed by pointer, and never by value. However, there are two places in `secp256k1_modinv64_var` where structure assignment is used. To get around this limitation we had to deviate slightly from the original code, and replace these two assignments with a subroutine that implements the assignment by copying the structure field by field.

## 9 Future Work

We chose to verify the modular inverse function because it is one of the more complicated algorithms implement in libsecp256k1. Because we are able to prove the correctness of this function, there is no reason to believe we couldn't verify the entirety of libsecp256k1 in the same way.

This formalization effort is part of the Simplicity[13] project to create an alternative language to Bitcoin script. Ultimately we would like to prove the implementations of all of Simplicity's intrinsics, known as jets, correctly implement their formal specifications. Again, the cryptographic jets are among the most sophisticated jets defined. So if we are able to formally verify the correctness of these jets, we ought to be able to formally verify any other jet in the same way.

A lot of formalization effort was spent reasoning about mathematical equalities and inequalities involving expressions with division and modulo operations by constants (see Figure 8). Much of that work ought to have been able to be handled automatically by Coq's linear integer arithmetic tactic (`lia`). And indeed after we completed our proof, we discovered there is an optional feature for Coq's `lia` tactic to enable support for such expressions. The next step would be to turn on this feature to see how much we can reduce and our formal proof by.

# Appendix A  Proof that secp256k1_umul128 meets its specification

```
Lemma body_secp256k1_umul128: semax_body Vprog Gprog f_secp256k1_umul128 secp256k1_umul128_spec.
Proof.
start_function.
repeat progressC.
* f_equal.
  apply Int64.eqm_repr_eq.
  eapply Int64.eqm_trans;[apply Int64.eqm_unsigned_repr|apply Int64.eqm_refl2].
  rewrite !Int64.unsigned_repr_eq.
  change Int64.modulus with (2^64).
  rewrite !Z.shiftr_div_pow2 by lia.
  match goal with
   |- (?x mod 2^64) = _ => replace x with
  (2 ^ 32 * ((a mod 2 ^ 32 * (b / 2 ^ 32)) mod 2 ^ 32) +
  2 ^ 32 * ((a / 2 ^ 32 * (b mod 2 ^ 32)) mod 2 ^ 32)  +
  (2 ^ 32 * (a mod 2 ^ 32 * (b mod 2 ^ 32) / 2 ^ 32) +
  (a mod 2 ^ 32 * (b mod 2 ^ 32)) mod 2 ^ 32)) by ring
  end.
  rewrite <- Z_div_mod_eq by lia.
  rewrite <- !(Zmult_mod_distr_l _ _ (2^32)).
  rewrite <- (Z.mod_small (a mod 2 ^ 32 * (b mod 2 ^ 32)) (2^64)) by (rewrite strict_bounds; solve_bounds).
  change (2^32 * 2^32) with (2^64).
  rewrite Zplus_mod_idemp_r, <- Zplus_assoc, Zplus_mod_idemp_l.
  match goal with
   |- (?x mod 2^64) = _ => replace x with
  ((2 ^ 32 * (a / 2 ^ 32 * (b mod 2 ^ 32)) mod 2^64 +
  (a mod 2 ^ 32 * (2 ^ 32 * (b / 2 ^ 32) + b mod 2 ^ 32)))) by ring
  end.
  rewrite <- Z_div_mod_eq by lia.
  change (2^64) with (2^32 * 2^32) at 1.
  rewrite Zmult_mod_distr_l, Zmult_mod_idemp_r, <- Zmult_mod_distr_l, Zplus_mod_idemp_l.
  match goal with
   |- (?x mod 2^64) = _ => replace x with
  ((2 ^ 32 * (a / 2 ^ 32) + (a mod 2 ^ 32)) * b) by ring
  end.
  rewrite <- Z_div_mod_eq by lia.
  reflexivity.
* rewrite <- (Z.shiftr_shiftr _ 32 32), !Z.shiftr_div_pow2 by lia.
  rewrite <- Z.div_add_l by lia.
  match goal with
   |- context [Int64.repr (?x / 2^32)] => replace x with
  (((2 ^ 32 * (a mod 2 ^ 32 * (b / 2 ^ 32) / 2 ^ 32) + (a mod 2 ^ 32 * (b / 2 ^ 32)) mod 2 ^ 32) +
    (a mod 2 ^ 32 * (b mod 2 ^ 32) / 2 ^ 32)) +
   (a / 2 ^ 32 * (2 ^ 32 * (b / 2 ^ 32))) +
   (2 ^ 32 * (a / 2 ^ 32 * (b mod 2 ^ 32) / 2 ^ 32) + (a / 2 ^ 32 * (b mod 2 ^ 32)) mod 2 ^ 32))
   ) by ring
  end.
  rewrite <- !Z_div_mod_eq by lia.
  rewrite <- Z.mul_add_distr_l.
  rewrite <- Z_div_mod_eq by lia.
  rewrite <- Z.div_add_l by lia.
  rewrite <- Z.mul_assoc.
  rewrite <- Z.mul_add_distr_l.
  rewrite (Z.mul_comm _ (2^32)).
  rewrite <- Z_div_mod_eq by lia.
  rewrite Z.add_comm.
  rewrite <- Z.div_add_l by lia.
  rewrite (Z.mul_comm _ (2^32)), Z.mul_assoc.
  rewrite <- Z.mul_add_distr_r.
  rewrite <- Z_div_mod_eq by lia.
  entailer!.
Qed.
```

**Figure 12.** Transcript of interactive proof steps for proving the implementation of `secp256k1_umul128` satisfies the specification given by `secp256k1_umul128_spec` given in Figure 5.

# Appendix B  Functions from libsecp256k1's `modinv64_impl.h`

```
/* Compute the transition matrix and eta for 62 divsteps (variable time, eta=-delta).
 *
 * Input:   eta: initial eta
 *          f0:  bottom limb of initial f
 *          g0:  bottom limb of initial g
 * Output: t: transition matrix
 * Return: final eta
 *
 * Implements the divsteps_n_matrix_var function from the explanation.
 */
static int64_t secp256k1_modinv64_divsteps_62_var(int64_t eta, uint64_t f0, uint64_t g0, secp256k1_modinv64_trans2x2 *t) {
    /* Transformation matrix; see comments in secp256k1_modinv64_divsteps_62. */
    uint64_t u = 1, v = 0, q = 0, r = 1;
    uint64_t f = f0, g = g0, m;
    uint32_t w;
    int i = 62, limit, zeros;

    for (;;) {
        /* Use a sentinel bit to count zeros only up to i. */
        zeros = secp256k1_ctz64_var(g | (UINT64_MAX << i));
        /* Perform zeros divsteps at once; they all just divide g by two. */
        g >>= zeros;
        u <<= zeros;
        v <<= zeros;
        eta -= zeros;
        i -= zeros;
        /* We're done once we've done 62 divsteps. */
        if (i == 0) break;
        VERIFY_CHECK((f & 1) == 1);
        VERIFY_CHECK((g & 1) == 1);
        VERIFY_CHECK((u * f0 + v * g0) == f << (62 - i));
        VERIFY_CHECK((q * f0 + r * g0) == g << (62 - i));
        /* Bounds on eta that follow from the bounds on iteration count (max 12*62 divsteps). */
        VERIFY_CHECK(eta >= -745 && eta <= 745);
        /* If eta is negative, negate it and replace f,g with g,-f. */
        if (eta < 0) {
            uint64_t tmp;
            eta = -eta;
            tmp = f; f = g; g = -tmp;
            tmp = u; u = q; q = -tmp;
            tmp = v; v = r; r = -tmp;
            /* Use a formula to cancel out up to 6 bits of g. Also, no more than i can be cancelled
             * out (as we'd be done before that point), and no more than eta+1 can be done as its
             * sign will flip again once that happens. */
            limit = ((int)eta + 1) > i ? i : ((int)eta + 1);
            VERIFY_CHECK(limit > 0 && limit <= 62);
            /* m is a mask for the bottom min(limit, 6) bits. */
            m = (UINT64_MAX >> (64 - limit)) & 63U;
            /* Find what multiple of f must be added to g to cancel its bottom min(limit, 6)
             * bits. */
            w = (f * g * (f * f - 2)) & m;
        } else {
            /* In this branch, use a simpler formula that only lets us cancel up to 4 bits of g, as
             * eta tends to be smaller here. */
            limit = ((int)eta + 1) > i ? i : ((int)eta + 1);
            VERIFY_CHECK(limit > 0 && limit <= 62);
            /* m is a mask for the bottom min(limit, 4) bits. */
            m = (UINT64_MAX >> (64 - limit)) & 15U;
            /* Find what multiple of f must be added to g to cancel its bottom min(limit, 4)
             * bits. */
            w = f + (((f + 1) & 4) << 1);
            w = (-w * g) & m;
        }
        g += f * w;
        q += u * w;
        r += v * w;
        VERIFY_CHECK((g & m) == 0);
    }
    /* Return data in t and return value. */
    t->u = (int64_t)u;
    t->v = (int64_t)v;
    t->q = (int64_t)q;
    t->r = (int64_t)r;
#ifdef VERIFY
    /* The determinant of t must be a power of two. This guarantees that multiplication with t
     * does not change the gcd of f and g, apart from adding a power-of-2 factor to it (which
     * will be divided out again). As each divstep's individual matrix has determinant 2, the
     * aggregate of 62 of them will have determinant 2^62. */
    VERIFY_CHECK(secp256k1_modinv64_det_check_pow2(t, 62, 0));
#endif
    return eta;
}
```

**Figure 13.** Source code of libsecp256k1's `secp256k1_modinv64_divsteps_62_var` function which is one of the functions we verified correct.

```c
/* Compute the inverse of x modulo modinfo->modulus, and replace x with it (variable time). */
static void secp256k1_modinv64_var(secp256k1_modinv64_signed62 *x, const secp256k1_modinv64_modinfo *modinfo) {
    /* Start with d=0, e=1, f=modulus, g=x, eta=-1. */
    secp256k1_modinv64_signed62 d = {{0, 0, 0, 0, 0}};
    secp256k1_modinv64_signed62 e = {{1, 0, 0, 0, 0}};
    secp256k1_modinv64_signed62 f, g;
    secp256k1_modinv64_signed62_assign(&(f), &(modinfo->modulus));
    secp256k1_modinv64_signed62_assign(&(g), &(*x));

#ifdef VERIFY
    int i = 0;
#endif
    int j, len = 5;
    int64_t eta = -1; /* eta = -delta; delta is initially 1 */
    int64_t cond, fn, gn;

    /* Do iterations of 62 divsteps each until g=0. */
    while (1) {
        /* Compute transition matrix and new eta after 62 divsteps. */
        secp256k1_modinv64_trans2x2 t;
        eta = secp256k1_modinv64_divsteps_62_var(eta, f.v[0], g.v[0], &t);
        /* Update d,e using that transition matrix. */
        secp256k1_modinv64_update_de_62(&d, &e, &t, modinfo);
        /* Update f,g using that transition matrix. */
#ifdef VERIFY
        VERIFY_CHECK(secp256k1_modinv64_mul_cmp_62(&f, len, &modinfo->modulus, -1) > 0); /* f > -modulus */
        VERIFY_CHECK(secp256k1_modinv64_mul_cmp_62(&f, len, &modinfo->modulus, 1) <= 0); /* f <= modulus */
        VERIFY_CHECK(secp256k1_modinv64_mul_cmp_62(&g, len, &modinfo->modulus, -1) > 0); /* g > -modulus */
        VERIFY_CHECK(secp256k1_modinv64_mul_cmp_62(&g, len, &modinfo->modulus, 1) < 0);  /* g <  modulus */
#endif
        secp256k1_modinv64_update_fg_62_var(len, &f, &g, &t);
        /* If the bottom limb of g is zero, there is a chance that g=0. */
        if (g.v[0] == 0) {
            cond = 0;
            /* Check if the other limbs are also 0. */
            for (j = 1; j < len; ++j) {
                cond |= g.v[j];
            }
            /* If so, we're done. */
            if (cond == 0) break;
        }

        /* Determine if len>1 and limb (len-1) of both f and g is 0 or -1. */
        fn = f.v[len - 1];
        gn = g.v[len - 1];
        cond = ((int64_t)len - 2) >> 63;
        cond |= fn ^ (fn >> 63);
        cond |= gn ^ (gn >> 63);
        /* If so, reduce length, propagating the sign of f and g's top limb into the one below. */
        if (cond == 0) {
            f.v[len - 2] |= (uint64_t)fn << 62;
            g.v[len - 2] |= (uint64_t)gn << 62;
            --len;
        }
#ifdef VERIFY
        VERIFY_CHECK(++i < 12); /* We should never need more than 12*62 = 744 divsteps */
        VERIFY_CHECK(secp256k1_modinv64_mul_cmp_62(&f, len, &modinfo->modulus, -1) > 0); /* f > -modulus */
        VERIFY_CHECK(secp256k1_modinv64_mul_cmp_62(&f, len, &modinfo->modulus, 1) <= 0); /* f <= modulus */
        VERIFY_CHECK(secp256k1_modinv64_mul_cmp_62(&g, len, &modinfo->modulus, -1) > 0); /* g > -modulus */
        VERIFY_CHECK(secp256k1_modinv64_mul_cmp_62(&g, len, &modinfo->modulus, 1) < 0);  /* g <  modulus */
#endif
    }

    /* At this point g is 0 and (if g was not originally 0) f must now equal +/- GCD of
     * the initial f, g values i.e. +/- 1, and d now contains +/- the modular inverse. */
#ifdef VERIFY
    /* g == 0 */
    VERIFY_CHECK(secp256k1_modinv64_mul_cmp_62(&g, len, &SECP256K1_SIGNED62_ONE, 0) == 0);
    /* |f| == 1, or (x == 0 and d == 0 and |f|=modulus) */
    VERIFY_CHECK(secp256k1_modinv64_mul_cmp_62(&f, len, &SECP256K1_SIGNED62_ONE, -1) == 0 ||
                 secp256k1_modinv64_mul_cmp_62(&f, len, &SECP256K1_SIGNED62_ONE, 1) == 0 ||
                 (secp256k1_modinv64_mul_cmp_62(x, 5, &SECP256K1_SIGNED62_ONE, 0) == 0 &&
                  secp256k1_modinv64_mul_cmp_62(&d, 5, &SECP256K1_SIGNED62_ONE, 0) == 0 &&
                   (secp256k1_modinv64_mul_cmp_62(&f, len, &modinfo->modulus, 1) == 0 ||
                    secp256k1_modinv64_mul_cmp_62(&f, len, &modinfo->modulus, -1) == 0)));
#endif

    /* Optionally negate d, normalize to [0,modulus), and return it. */
    secp256k1_modinv64_normalize_62(&d, f.v[len - 1], modinfo);
    secp256k1_modinv64_signed62_assign(&(*x), &(d));
}
```

**Figure 14.** Source code of libsecp256k1's `secp256k1_modinv64_var` function whose correctness proof we completed was the ultimate goal of this verification effort.